# Numerical analysis on the spatiotemporal characteristics of the Portevin-Le Chatelier effect in Ti-12Mo alloy

Shiyuan Luo[a, b, *], Yongxin Jiang[a, b], Sandrine Thuillier[c], Philippe Castany[d], Liangcai Zeng[a, e]

[a]*Key Laboratory of Metallurgical Equipment and Control Technology, Ministry of Education, Wuhan University of Science and Technology, Wuhan 430081, China*

[b]*Hubei Key Laboratory of Mechanical Transmission and Manufacturing Engineering, Wuhan University of Science and Technology, Wuhan 430081, China*

[c]*Univ. Bretagne Sud, UMR CNRS 6027, IRDL, F-56100 Lorient, France*

[d]*Univ Rennes, INSA Rennes, CNRS, ISCR-UMR 6226, F-35000 Rennes, France*

[e]*Precision Manufacturing Institute, Wuhan University of Science and Technology, Wuhan 430081, China*

**Abstract**

A simplified 3D FE model based on McCormick's model is developed to numerically predict the spatiotemporal behaviors of the PLC effect in Ti-12Mo alloy tensile tests at 350 °C with an applied strain rate of the order of $10^{-3}$ s$^{-1}$. The material parameter identification procedure is firstly presented in details, and the simulated results are highly consistent with experimental ones, especially in terms of stress drop magnitudes and PLC band widths. Specifically, the distribution of simulated stress drop magnitudes follows a normal distribution and its peak value is in the range of 26-28 MPa. The simulated band width slightly fluctuates with the increase of true strain and its average value is about 1.5 mm. Besides, the staircase behavior of strain-time curves and the hopping propagation of the PLC band are observed in Ti-12Mo alloy tensile process, which are related to the strain localization and stress drop magnitudes.

---

* Corresponding author. Tel.: +86 27 68862283. E-mail address: shiyuanluo@wust.edu.cn (S. Luo)

# 1. Introduction

Titanium alloys are extensively used in numerous areas from aeronautics to biomedical devices, due to their high strength-to-weight ratio, good corrosion resistance and excellent biocompatibility [1-3]. Hereinto, metastable β Ti-xMo alloys synthesized by nontoxic elements are regarded as ones of promising metallic orthopedic implant materials [4-7]. However, according to previous studies, a type of plastic instability, namely Portevin-Le Chatelier (PLC) effect, is observed in Ti-xMo alloys under various thermo-mechanical loading conditions [8-10]. Therefore, to accurately predict the spatiotemporal behaviors of the PLC effect in Ti-xMo alloys is of great importance for the process optimization and practical application of these materials.

The PLC effect exhibits itself as serrated flow in stress-strain curves and localized deformation bands in specimens of metallic alloys [11]. Until now, some researches have been performed on the PLC effect from temporal and spatial aspects. Focusing on the former, Chihab et al. [12] and Amokhtar et al. [13] respectively investigated the influence of strain rates and alloying element contents on stress drop magnitudes in Al-Mg alloys, and experimental results indicated that the stress drop magnitude is negative correlation with strain rates and positive correlation with Mg contents. Moreover, Zhang et al. [14] studied stress drop magnitudes in Ni-Co superalloy tensile tests within the temperature range of 350-550 °C, and found that the stress drop magnitude increases when the temperature rises. Furthermore, Chen et al. [15] studied the effects of temperature and strain rate on the critical strain of the PLC effect in HfNbTaTiZr high-entropy alloy, and discovered that unlike the normal behavior of strain rate effect, the evolution trend of critical strain transfers from a normal behavior to an inverse behavior as enhancing the temperature to a given value. On the other hand, for the spatial aspect, the PLC band propagation is classified as types A, B and C, which are respectively continuous propagation, hopping propagation and random nucleation [16-18]. Mehenni et al. [19] compared the propagation characteristics of the PLC bands at different strain rates in Al-Mg alloys, and the results suggested that the propagation type of the PLC band transfers from type C to type B and eventually to type A when increasing the strain rate. Yu et al. [20] reported that the inclined angles of PLC bands range within 50°-60° with respect to the tensile direction during the propagation process. Besides, some studies [21-23] were performed on quantitatively characterizing the variation of the PLC band width at different testing conditions. Note that, all of the aforementioned investigations are performed by experiments rather than finite element (FE) methods.

Focusing on the modeling of the PLC effect, a mathematical model is firstly developed by Kubin and Estrin [24], which is based on a macroscopic description of deformation bands. However, this model has difficulties in quantitatively describing the spatiotemporal characteristics of the PLC effect. Hence, according to a microscopic description of the dynamic strain ageing, McCormick [25] proposed another mathematical model, in which the negative strain rate sensitivity and serrations are implicit consequences of constitutive equations. Then, Zhang et al. [26] implemented McCormick's model in the Abaqus software as an external subroutine, and simulated the PLC effects of both flat and round specimens in Al-Mg-Si alloy tensile tests. Moreover, Mazière et al. [27] employed McCormick's model to analyze the temporal behavior of the PLC effect in nickel-based superalloy. Furthermore, Manach et al. [28] used McCormick's model to analyze the PLC effect in an Al-Mg alloy, and compared the simulated results with experimental ones in terms of stress drop magnitudes and propagation types of the PLC bands. Mansouri et al. [29] employed McCormick's model to predict the PLC effect of aluminium alloys in the case of room temperature Erichsen tests. Besides, considering the temperature effect, Belotteau et al. [30] improved McCormick's model by using temperature dependent material parameters. Mansouri et al. [31] and Moon et al. [32] utilized the improved McCormick's model to predict the PLC effect of aluminum alloys during tensile and simple shear as well as limiting dome height formability tests. Note that, although a large number of simulation works have been performed on the PLC effect in various metallic alloys, however, none of them is tied to the prediction of the PLC effect in Ti-Mo alloy.

Consequently, this paper is aimed to utilize FE methods to investigate the spatiotemporal behaviors of the PLC effect in Ti-12Mo alloy. For this purpose, McCormick's model is employed and its material parameters are identified in details. Then, a simplified 3D FE model based on the calibrated McCormick's model is developed by using Abaqus software and verified by experiments. Finally, the spatiotemporal characteristics of the PLC effect in Ti-12Mo alloy tensile tests at 350 °C with an applied strain rate of the order of $10^{-3}$ s$^{-1}$ are numerically analyzed.

2. **Experimental procedures**

Ti-12Mo alloy is synthesized in a water-cooled Cu crucible by using cold crucible levitation melting technique with pure titanium (99.95 wt.%) and molybdenum (99.99 wt.%). Then, to ensure the uniformity of composition and remove the surface oxide layers of ingots, a homogenization treatment at 950 °C for 20 hours followed by water quench and acidic bath (50 vol.% HF and 50 vol.% HNO$_3$) is performed. After that, the ingots are cold-rolled into plates with a 90% thickness reduction, leading to a final thickness around 1 mm. Subsequently, tensile test specimens with a gauge size of 10×3×1mm$^3$ are cut from these rolled plates along the rolling direction, as shown in Fig. 1(a). Furthermore, to analyze the deformation characteristics of specimens during tensile tests, a black and white contrast pattern is sprayed over the specimen surface prior to tests, as presented in the partial enlarged drawing of Fig.

1(b). Finally, using Gleeble 3500 testing machine coupled with a digital image correlation technique, Ti-12Mo alloy tensile tests are performed at 350 °C with three different applied strain rates from the order of $10^{-4}$ s$^{-1}$ to $10^{-2}$ s$^{-1}$. Note that, during the whole testing process, the maximum temperature gradient along the tensile direction measured by three thermocouples is less than 8.5% (Fig. 1(c)), and the temperature fluctuation of each thermocouple is maintained within 2 °C. Therefore, the experimental tests can be regarded as isothermal tensile tests within the gauge size. Additionally, the corresponding engineering stress-strain curves are calculated and converted to the true stress-strain curves, as shown in Fig. 1(d).

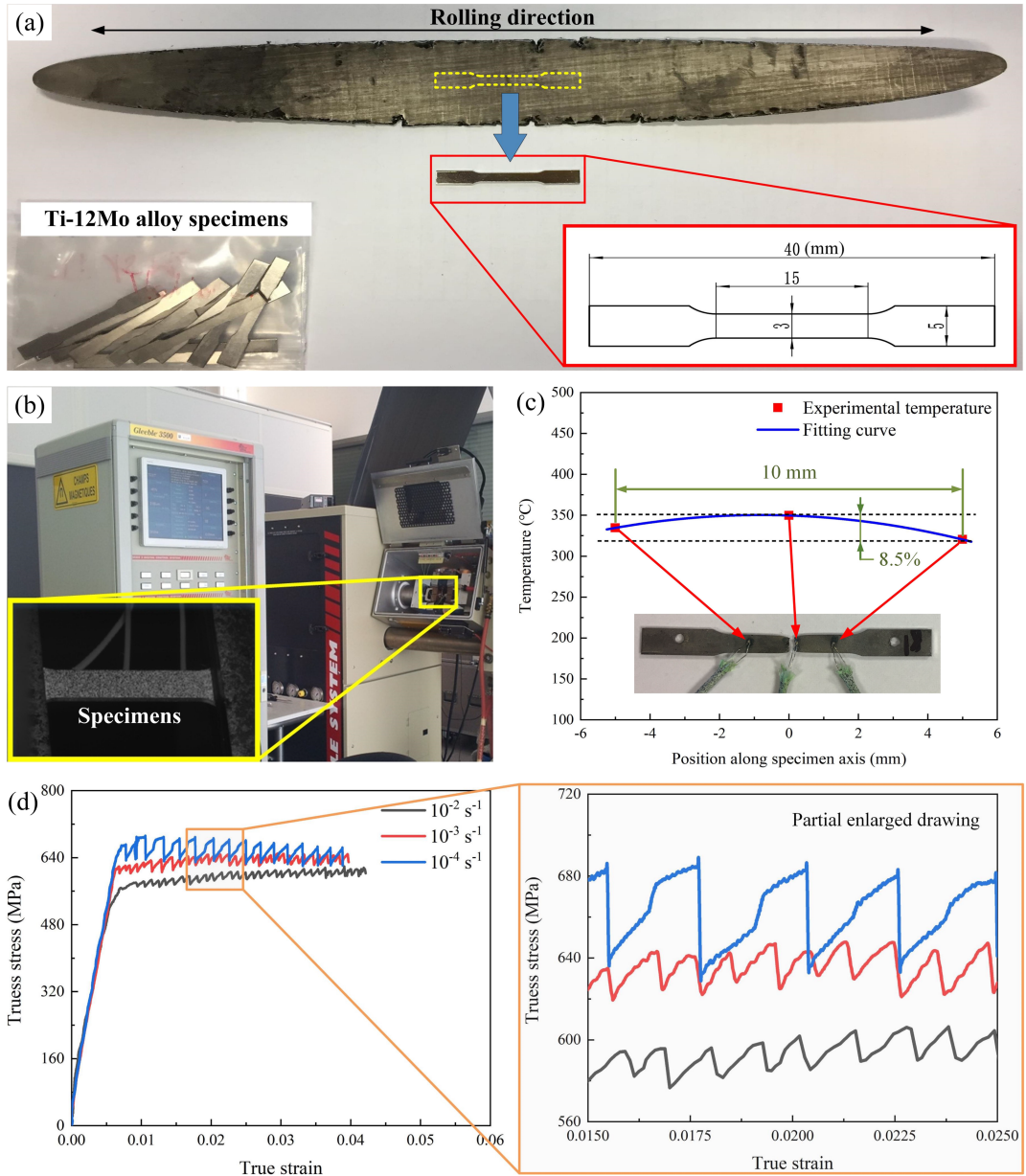

**Fig. 1.** Ti-12Mo alloy specimens (a), experimental set-up (b), temperature gradient (c) and true stress-strain curves with partial enlarged drawing (d).

## 3. Modeling of the PLC effect

*3.1 Constitutive model*

Based on an elasto-visco-plastic frame work, McCormick's model [33-35] is employed to describe the PLC effect in Ti-12Mo alloy. Hereinto, the total deformation tensor is the sum of elastic and plastic strain tensors, as shown below:

$$\underset{\sim}{\varepsilon} = \underset{\sim}{\varepsilon}^e + \underset{\sim}{\varepsilon}^p, \quad \underset{\sim}{\sigma} = \underset{\approx}{c} : \underset{\sim}{\varepsilon}^e \tag{1}$$

Where $\underset{\sim}{\varepsilon}^e$ and $\underset{\sim}{\varepsilon}^p$ are respectively the elastic and plastic strain tensors, $\underset{\sim}{\sigma}$ is the true stress tensor, and $\underset{\approx}{c}$ is the fourth-rank tensor of elastic modulus.

Moreover, the yield function is defined as:

$$f(\underset{\sim}{\sigma}, p, t_a) = \sigma_{eq} - R(p) - R_a(p, t_a), \quad \sigma_{eq} = \sqrt{\frac{3}{2} \underset{\sim}{s} : \underset{\sim}{s}} \tag{2}$$

where $\sigma_{eq}$ denotes the equivalent stress, $R(p)$ means the strain hardening, $R_a(p, t_a)$ is the yield stress, $p$ is the equivalent plastic strain, $t_a$ is an internal variable considering the ageing contribution to the yield stress, and $\underset{\sim}{s}$ is the deviatoric part of the stress tensor. Hereinto, $R(p)$ can be defined as:

$$R(p) = Q(1 - \exp(-bp)) \tag{3}$$

where $Q$ and $b$ are hardening parameters.

Furthermore, the yield stress $R_a(p, t_a)$ is written as:

$$R_a(p, t_a) = R_0 + P_1 C_m \left(1 - \exp\left(-P_2 p^\alpha t_a^n\right)\right) \tag{4}$$

where $R_0$ is related to the initial yield stress, $P_1 C_m$ corresponds to the stress drop magnitude, $P_2$, $\alpha$ and $n$ are related to the saturation rate, and the evolution equation of ageing time $t_a$ is shown as below:

$$\dot{t}_a = \frac{t_w - t_a}{t_w}, \quad t_w = \frac{w}{\dot{p}} \tag{5}$$

where $t_w$ denotes the mean waiting time of dislocations, $w$ is the increment of plastic strain generated when all the stopped dislocations overcome the obstacles, and $\dot{p}$ is the equivalent plastic strain rate calculated by the following viscoplastic hyperbolic flow rule.

$$\dot{p} = \dot{p}_0 \sinh\left(\frac{\langle f(\underset{\sim}{\sigma}, p, t_a) \rangle}{K}\right) \tag{6}$$

where the brackets $\langle f(\underset{\sim}{\sigma}, p, t_a) \rangle$ means the maximum value between $f(\underset{\sim}{\sigma}, p, t_a)$ and 0, $\dot{p}_0$ and $K$ are material parameters.

Note that, prior to the critical plastic strain of the PLC effect, the equivalent plastic strain rate $\dot{p}$ is nearly constant as increasing the plastic deformation. Therefore, the explicit expression of $t_a$ in Eq. (5) can be expressed as a function of $p$ [27]:

$$t_a(p) = \frac{\omega}{\dot{p}}\left(1-\exp\left(-\frac{p}{\omega}\right)\right) + \frac{R_0}{E\dot{p}}\exp\left(-\frac{p}{\omega}\right) \quad (7)$$

Finally, coupling above equations, the homogeneous solution of stress σ can be written as:

$$\sigma = K\operatorname{arcsinh}\left(\frac{\dot{p}}{\dot{p}_0}\right) + R_0 + Q(1-\exp(-bp)) + P_1 C_m\left(1-\exp\left(-P_2 p^\alpha t_a(p)^n\right)\right) \quad (8)$$

*3.2 Material parameter identification*

In above equations, 13 material parameters need to be determined considering elastic and plastic deformation. The identification methods are presented in details below.

For elastic part of the strain in Ti-12Mo alloy, it is regarded as linear variation. Young's Modulus ($E$) can be obtained by linear fitting the experimental true stress-strain curve in the stress range [100-400 MPa], as shown in Fig. 2(a). The result indicates that the linear fitting curve shows a good agreement with the experimental one (Fig. 2(b)), and the value of $E$ is 92.890 GPa. Moreover, with the same tensile testing conditions, the value of $E$ reported by literature [8] is about 92 GPa, which further verifies the validity and accuracy of the fitting result. Besides, Poisson's ratio ($v$) of Ti-12Mo alloy is 0.33, referring to [36].

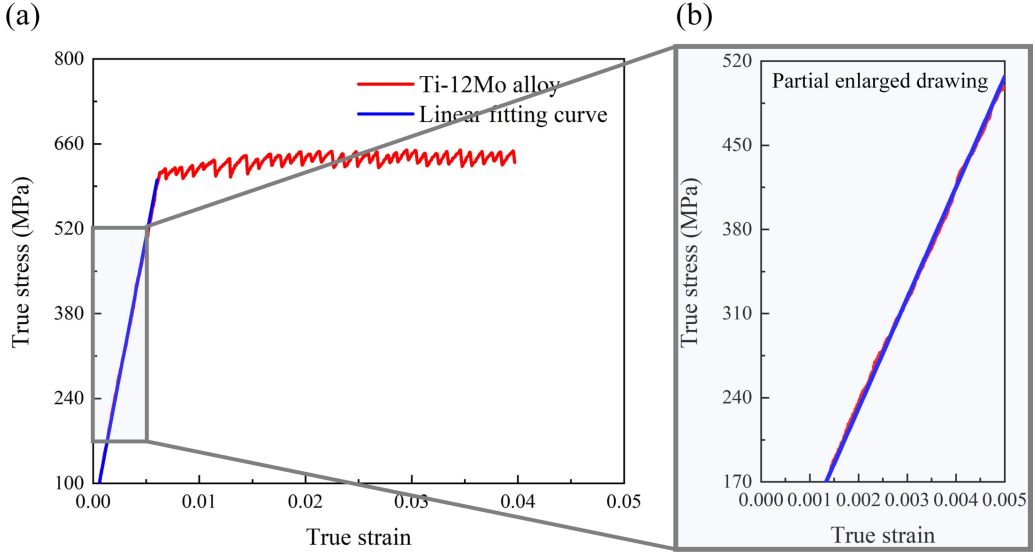

**Fig. 2.** Identification method for Young's Modulus of Ti-12Mo alloy (a) with partial enlarged drawing (b).

Then, for the plastic part of Ti-12Mo alloy, the true stress-plastic strain curve of Ti-12Mo alloy is fitted by $\sigma=\sigma_y+Q(1-\exp(-bp))$ with Levenberg Marquardt iteration algorithm [31], as presented in Fig. 3(a). It can be seen that the smooth curve effectively describes the hardening trend of the experimental curve, and the fitting parameters $\sigma_y$, $Q$ and $b$ are respectively 607.854 MPa, 27.075 MPa and 220.458. Moreover, according to literature [30, 35], three parameters ($n$, $w$ and $C_m$) in Eqs. (4) and (5) are determined, whose values respectively are 0.66, 0.0001 and 1. Furthermore, in order to identify $R_0$, viscous ($K$, $\dot{p}_0$) and ageing ($P_1$, $P_2$, $\alpha$) parameters, tensile tests at 350 °C with three different applied strain rates from the order of $10^{-4}$ s$^{-1}$ to $10^{-2}$ s$^{-1}$ are performed (Fig. 1(d)). Thereafter, from experimental results, true stress values at $p=0.02$ under each strain rates are measured and depicted as red points in Fig. 3(b). Subsequently, using the same iteration method as mentioned above, these points are fitted by Eq. (8). It can be seen that the nonlinear fitting curve is highly consistent with experimental data, which can prove the validity of the fitting results. Finally, the corresponding parameters ($R_0$, $K$, $\dot{p}_0$, $P_1$, $P_2$, $\alpha$) and aforementioned calibrated parameters ($E$, $v$, $Q$, $b$, $n$, $w$, $C_m$) are summarized and listed in Table 1.

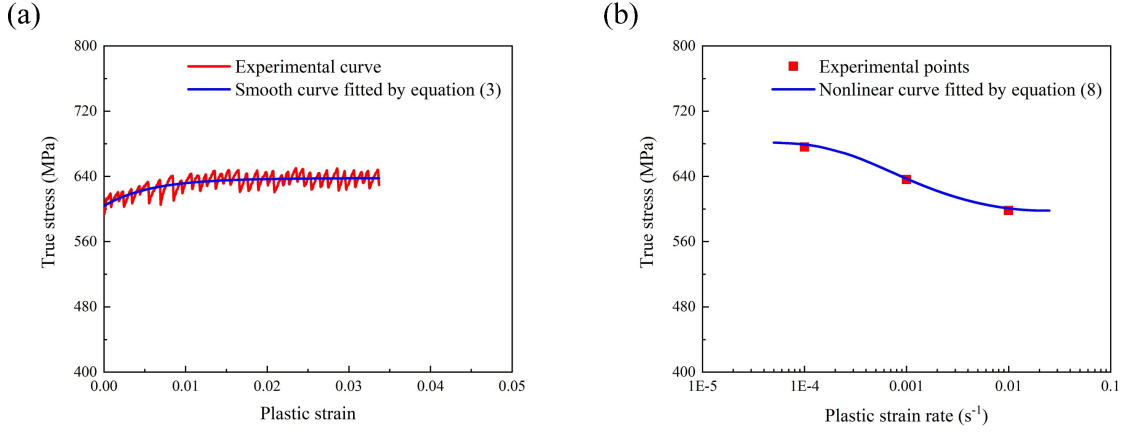

**Fig. 3.** Identification methods for $Q$, $b$ and $\sigma_y$ (a) as well as $K$, $\dot{p}_0$, $P_1$, $P_2$ and $\alpha$ (b) of Ti-12Mo alloy.

**Table 1** Material parameters for modeling the PLC effect in Ti-12Mo alloy.

| Parameters | Units | Values |
| --- | --- | --- |
| $E$ | GPa | 92.890 |
| $v$ | - | 0.33 [36] |
| $R_0$ | MPa | 555 |
| $Q$ | MPa | 27.075 |
| $b$ | - | 220.458 |
| $K$ | MPa | 15 |
| $\dot{p}_0$ | $s^{-1}$ | 0.05 |
| $P_1$ | MPa | 100 |
| $P_2$ | $s^{-n}$ | 15 |
| $C_m$ | at.% | 1 [35] |
| $\alpha$ | - | 0.36 |
| $w$ | - | $10^{-4}$ [35] |
| $n$ | - | 0.66 [30] |

*3.3 FE validation*

In contrast to explicit solution techniques, the implicit method can analyze static and quasi-static events easily and accurately, due to the reason that it is unconditionally stable with respect to the size of the time increment [37]. Therefore, to simulate the PLC effect in Ti-12Mo alloy, the above constitutive model with calibrated material parameters presented in Table 1 is implemented into Abaqus finite element software as an external subroutine, using an implicit scheme. Fig. 4(a) shows the established 3D FE model with geometric dimensions, boundary conditions and mesh details. Specifically, to simplify the model and improve computational efficiency, only the middle part of tensile test specimen with the size of 15×3×1 mm$^3$ is employed for 3D FE modeling. Moreover, to effectively simulate the tensile process, the displacement on left side of the FE model is restricted, and a constant velocity (0.01 mm/s) is applied to the right side. Besides, for mesh details, it should be pointed out that the element size is defined as 0.25×0.25 mm$^2$, and eight node solid elements with reduced integration and enhanced hourglass control (C3D8R) are used for the FE model to avoid zero energy modes. Fig. 4(b) compares the simulation results with experimental ones. It can be found that although the frequency of simulated stress drops is higher than the experimental one, the stress level is well predicted for both elastic and plastic regions, and errors on the value of average stress drop magnitude are less than 7.1%. Therefore, it can be concluded that the 3D FE model with the constitutive model and material parameters mentioned in Sections 3.1 and 3.2 is valid for simulating the PLC effect in Ti-12Mo alloy.

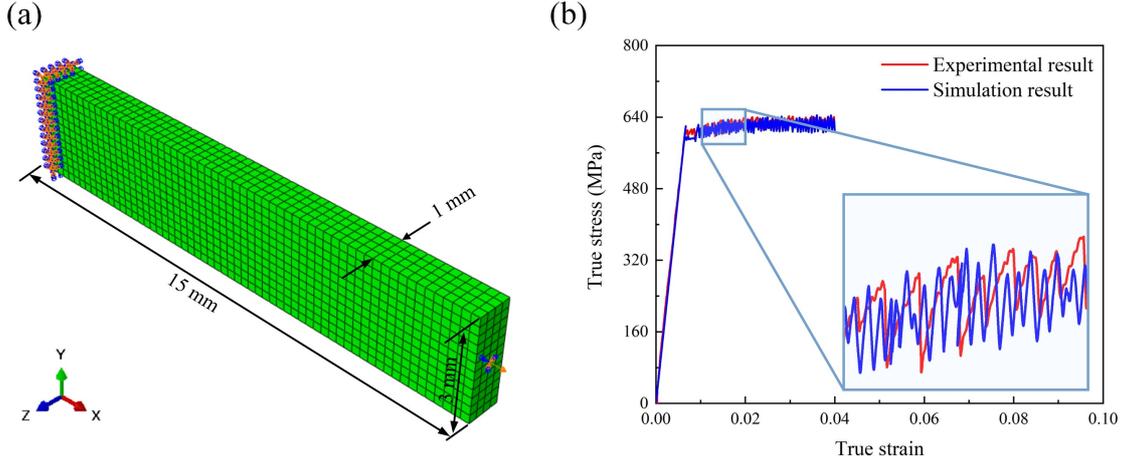

**Fig. 4.** 3D simplified FE model with geometric dimensions, boundary conditions and mesh details (a) as well as the comparison of simulated and experimental results (b).

## 4. Results and discussion

*4.1 Strain-time curve and strain localization*

Fig. 5(a) shows the variation of true strain versus the time at the constant simulated tensile velocity (0.01 mm/s). For each time, the true strain is calculated by $\varepsilon=\ln(1+\Delta L/L_1)$, where $\Delta L$ is the displacement difference of two points ($P_1$ and $P_3$) and $L_1$ is the gauge length (10 mm). It can be obtained from Fig. 5(a) that the slope of the true strain-time curve is about 0.00095 s$^{-1}$, which is in accordance with the experimental strain rate (about 0.0011 s$^{-1}$). Moreover, Fig. 5(b) presents the variation of $p$ with the increase of the time at three points ($P_1$, $P_2$ and $P_3$). It can be seen that the $p$ values of three points are 0 prior to 10 s. This is because this tensile process mainly results in elastic deformation within the tensile specimen. Furthermore, after 10 s, staircase behaviors are observed as increasing the tensile time, which is similar to the results reported by Mansouri et al. [31] and Benallal et al. [34]. This phenomenon is mainly attributed to the heterogeneous deformation during the tensile procedure. Finally, it can be found that the largest $p$ value at $P_2$ (about 0.045) is more than twice as big as the largest ones at $P_1$ and $P_3$, although the deformation beginning time of the former is later than the ones of the latter. This phenomenon can be explained by the reason that the localized plastic deformation firstly occurs at both ends of the tensile specimen, then gradually transfers to its middle part, and eventually stays at this place with the increase of the time, as shown in Fig. 5(c). Note that, to provide a maximum contrast, the color scale presented in Fig. 5(c) is independent for each frame, based on the instantaneous maximum and minimum values of equivalent plastic strain. The corresponding maximum equivalent plastic strain and time are indicated over and under each frame, respectively.

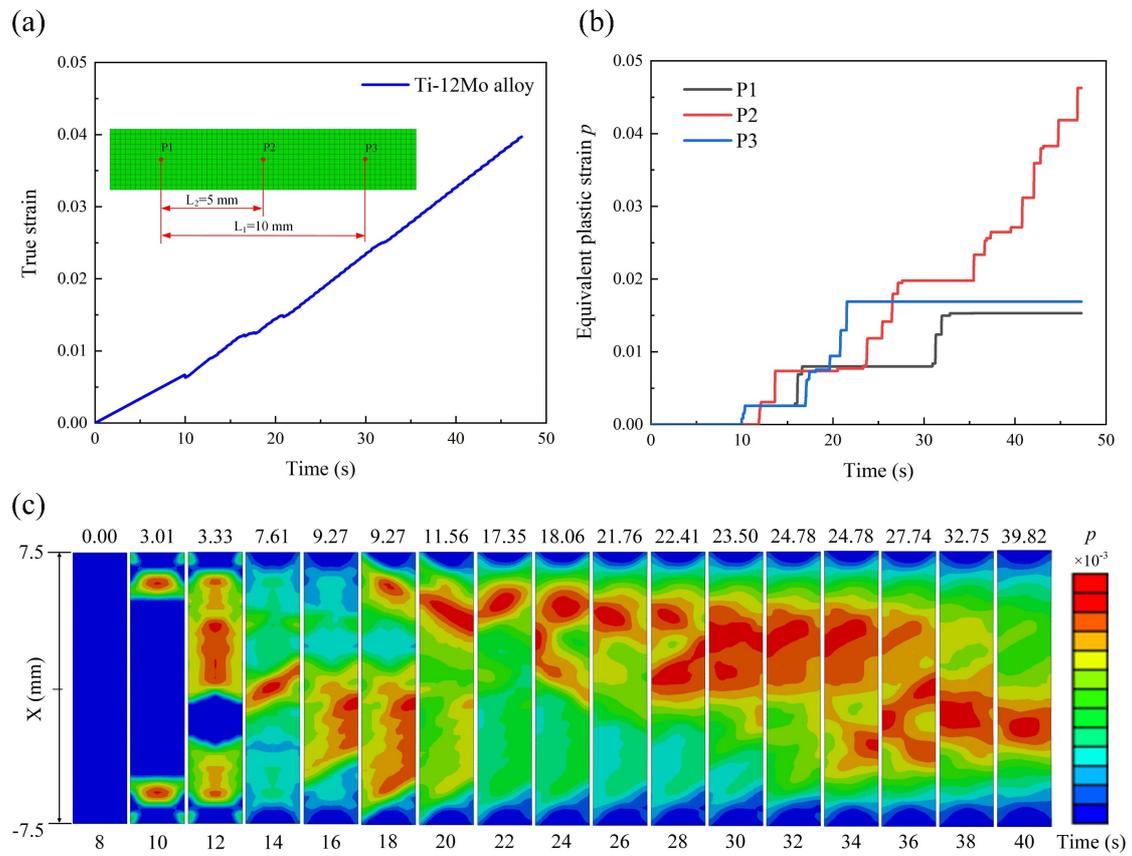

**Fig. 5.** True strain-time curve (a), equivalent plastic strain-time curves of three selected points (b), and the distribution of equivalent plastic strain in Ti-12Mo alloy simulated tensile procedure (c).

*4.2 Stress-time curve and PLC band propagation*

Fig. 6(a) shows the variation of true stress versus the time at the constant simulated tensile velocity (0.01 mm/s). For each time, the true stress is obtained by $\sigma=F(1+\Delta L/L_1)/A_0$, where $F$ is the tensile force and $A_0$ is the initial cross-sectional area of the tensile specimen (3 mm$^2$). It can be observed from Fig. 6(a) that the true stress shows serrated flow at the level of 620 MPa after a linear rise when the tensile time increases to the value of 10 s. This phenomenon is related to the transformation from elastic to plastic deformation. Moreover, in order to analyze the propagation characteristics of the PLC band, the partial enlarged drawing of the serrated part and the corresponding PLC band at the marked valley of the stress flow are depicted in Figs. 6(b) and (c), respectively. Like Fig. 5(c), the color scale presented in Fig. 6(c) is also independent for each frame with the corresponding maximum strain rate and time indicated over and under each frame. It can be observed from Fig. 6(c) that the hopping propagation of the PLC band appears with the increase of the tensile time, which indicates that a type B propagation will occur at 350 °C with the applied strain rate of the order of 10$^{-3}$ s$^{-1}$ in Ti-12Mo alloy tensile procedure. This phenomenon is closely consistent to the results reported by Luo et al. [9, 10], who investigated the propagation characteristics of the PLC bands in Ti-xMo alloys by using a digital image correlation method. Besides, like the angle variations of the PLC bands in an AlMg alloy reported by Yuzbekova et al [11], it can be found in Ti-12Mo alloy that the inclined angle of the PLC band also changes during the PLC band propagation process. This phenomenon is mainly related to variations of stress conditions influenced by stress drop magnitudes.

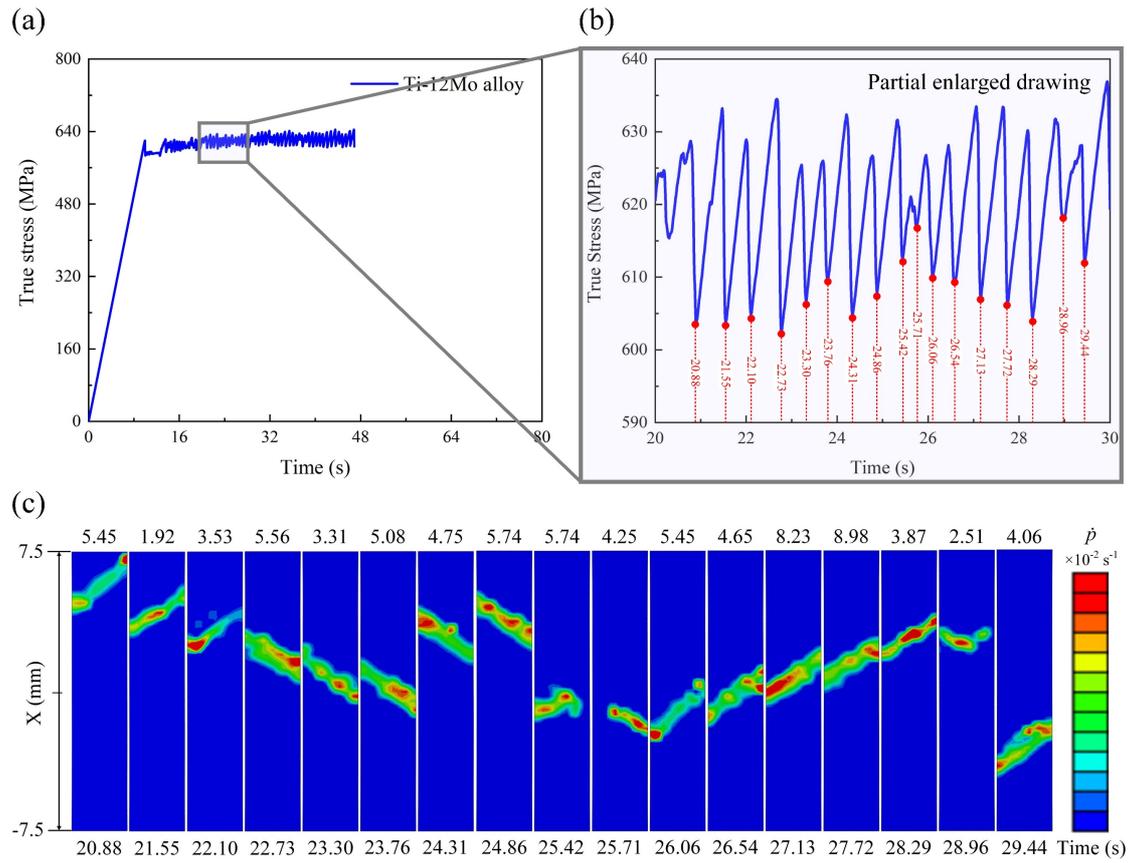

**Fig. 6.** True stress-time curve (a), partial enlarged drawing of selected area (b) and propagation of PLC bands in Ti-12Mo alloy simulated tensile procedure (c).

*4.3 Stress drop magnitude and PLC band width*

To quantitatively analyze the distribution of stress drop magnitudes in Ti-12Mo alloy tensile procedure, the variation of stress drop number versus the stress drop magnitude at the constant simulated tensile velocity (0.01 mm/s) is presented in Fig. 7(a). Hereinto, the stress drop magnitude is measured from simulated true stress-strain curve depicted in Fig. 4(b). It can be seen from Fig. 7(a) that the maximum and minimum values of stress drop magnitudes are about 4 MPa and 36 MPa, respectively. These values are slightly higher than experimental ones, whose maximum and minimum values are respectively about 2 MPa and 34 MPa. Moreover, the simulated and experimental values of average stress drop magnitudes are calculated by $\Delta\sigma_{average} = \sum_{1}^{N}(\Delta\sigma)/N$ [10], which are respectively 21.2 MPa and 19.7 MPa. Furthermore, it can be observed from the red trend line that the distribution of stress drop magnitudes follows a normal distribution and its peak value is in the range of 26-28 MPa. Besides, to quantitatively reflect strain localization characteristics, the variation of the PLC band width with the increase of true strain is described in Fig. 7(b). Note that, using the same measurement method reported by Sene et al. [38], the PLC band width is measured on the central axis of the specimen along the tensile direction. It can be observed from Fig. 7(b) that the value of the PLC band width slightly fluctuates with the increase of true strain, and its average value is about 1.5 mm. This phenomenon shows a good agreement with our previous experimental results (about 1.35 mm) [9].

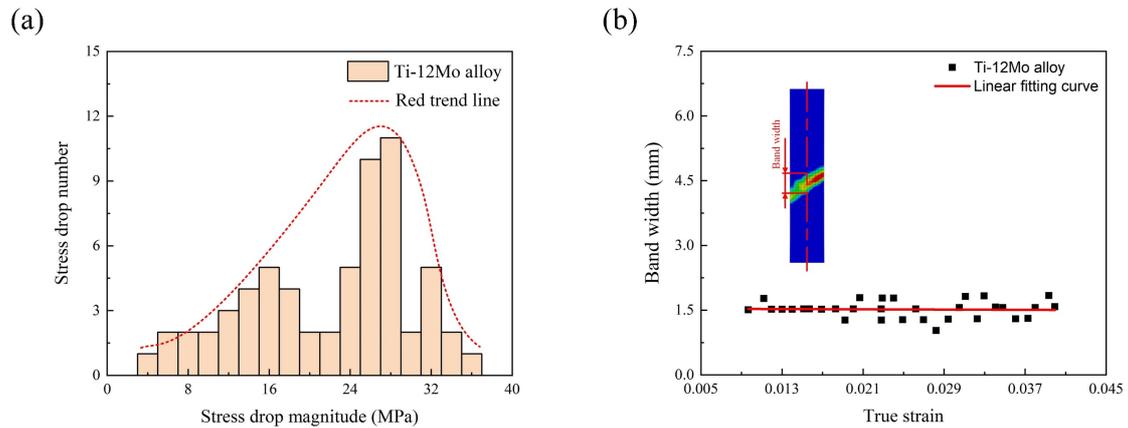

**Fig. 7.** Stress drop magnitude distribution (a) and PLC band width evolution (b) in Ti-12Mo alloy simulated tensile procedure.

5. **Conclusions**

In this paper, McCormick's model implemented in Abaqus software is used to simulate the spatiotemporal characteristics of the PLC effect in Ti-12Mo alloy tensile procedure and verified by experiments. The conclusions are summarized as following:

(1) McCormick's model with calibrated material parameters can well predict the spatiotemporal behaviors of the PLC effect in Ti-12Mo alloy tensile procedure, especially in terms of stress drop magnitudes and PLC band widths. Moreover, the identification methods of corresponding material parameters and the 3D FE modeling procedure are presented in details, which can provide a significant guidance for the simulation of PLC effect in other materials.

(2) For the temporal behaviors of the PLC effect in Ti-12Mo alloy, strain staircases are observed as increasing the tensile time, which can be attributed to the strain localization. Moreover, for its spatial behaviors, the hopping propagation of the PLC band appears and the inclined angle of the PLC band changes in Ti-12Mo alloy tensile procedure. These phenomena are mainly related to the variation of stress drop magnitudes.

(3) The distribution of simulated stress drop magnitudes follows a normal distribution and its peak value is in the range of 26-28 MPa. Moreover, the PLC band width slightly fluctuates with the increase of true strain and its average value is about 1.5 mm. Besides, to further numerically analyze the temperature and strain rate ranges for the occurrence of the PLC effect in Ti-12Mo alloy, the material parameters of McCormick's model will be identified as a function of temperature and the frequency of simulated stress drops will be optimized in the future work.

**Acknowledgements**

The authors would like to gratefully acknowledge the financial support from the Hubei Provincial Natural Science Foundation of China (nos. 2020CFB115).